\documentclass[a4paper,11pt]{article}
\usepackage{pos}
\title{Integration and qualification of the Mini-EUSO telescope on board the ISS }

\author*[a,e]{G. Cambi\'e}
\author[b,c]{A. Belov}
\author[d]{F. Capel}
\author[a,e,f]{M. Casolino}
\author[g]{A. Franceschi}
\author[c]{P. Klimov}
\author[a]{L.~Marcelli}
\author[f]{T. Napolitano}
\author[a,e]{P. Picozza}
\author[f]{L.W. Piotrowski}
\author[a]{E. Reali}
\author[g]{M. Ricci}

\affiliation[a]{Università degli studi di Roma Tor Vergata, Dipartimento di Fisica\\
  Roma, Italy}
\affiliation[b]{Faculty of Physics, M.V. Lomonosov Moscow State University\\
  Moscow, Russia}
\affiliation[c]{Skobeltsyn Institute of Nuclear Physics, Lomonosov Moscow State University\\
  Moscow, Russia}
\affiliation[d]{KTH Royal Institute of Technology\\
  Stockholm, Sweden}
\affiliation[e]{INFN, Sezione di Roma Tor Vergata \\
  Roma, Italy}
\affiliation[f]{Riken\\ 
  Wako, Japan}
\affiliation[g]{INFN-LNF\\
  Frascati, Italy}

\emailAdd{giorgio.cambie@roma2.infn.it}

\abstract{Mini-EUSO is a compact telescope ($37 \times 37 \times
62$~cm$^3$) currently hosted on board the International Space Station.
Mini-EUSO is devoted primarily to study Ultra High Energy Cosmic Rays
(UHECR) above $10^{21}$~eV but also to search for Strange Quark Matter
(SQM), to observe Transient Luminous Event (TLE) in upper atmosphere,
meteoroids, sea bioluminescence and space debris tracking. Mini-EUSO
consist of a main optical system, the Photo Detector Module (PDM),
sensitive to UV spectrum ($300\div400$~nm) and several ancillary sensors
comprising a visible ($400\div780$~nm) and NIR ($1500\div1600$~nm)
cameras and a $8 \times 8$ channels Multi-Pixel Photon Counter Silicon
PhotoMultiplier (MPPC SiPM) array which will increase the Tecnological
Readyness Level of this ultrafast imaging sensor.  Mini-EUSO belongs to
a novel set of missions committed to evaluate, for the first time, the
capability of observing Cosmic Rays from a space-based.  The
instrumentation, space-qualified tests will be shown.}

\FullConference{37$^{\rm{th}}$ International Cosmic Ray Conference (ICRC 2021)\\
		July 12th -- 23rd, 2021\\
		Online -- Berlin, Germany}

\begin{document}
\maketitle

\section{Introduction}

Mini-EUSO~\cite{Bacholle_2021} is a mission which belongs to the JEM-EUSO (Joint Experiment Missions - Extreme Universe Space Observatory) collaboration, which 
aim to study Ultra High Energy Cosmic Rays (UHECRs) from space for the first time. 
The collaboration has developed different former missions, which togheter with Mini-EUSO, are turning out to be a good test bench for the future, larger and better 
performing satellites, K-EUSO~\cite{2017ICRC...35..368C} 
(KLYPVE) and POEMMA~\cite{Olinto_2021} (Probe Of Extreme Multi-Messenger Astrophysics). 
Mini-EUSO was launched on August the 22nd 2019 with the Soyuz MS-14 spacecraft (an unmanned cargo expedition) from the Bajkonur Cosmodrome (Kazakhstan)
and now is  accommodated on the Russian Zvezda module, facing a UV-transparent window in Nadir mode.
Mini-EUSO is observing UV emissions looking at the Earth surface in the latitude range covered by the ISS ($\pm 51.6^{\circ}$).
This approach is unique, not only for the huge area covered respect to the onground experiments, 
but also because both Earth's hemispheres can be observed with one instrument, thus applying the same systematic error.

\section{The telescope}

\begin{figure}[h!]
 \centering
 \includegraphics[scale=1.3,keepaspectratio=true]{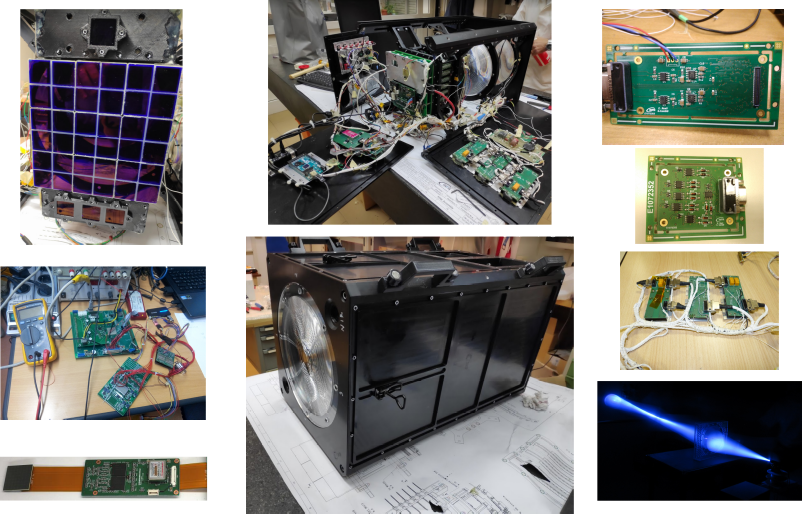}
 \caption{\label{int}Mini-EUSO hardware and integration.}
\end{figure}

The main detector has a super-wide-field of view ($44^{\circ}$) which allows to map 
an Earth ground area of $263 \times 263 \ km^2$ thanks to the optics which comprises two double sided, $25$ cm diameter, PMMA Fresnel lenses which focuses light onto a 
36 Hamamatsu Multi-Anode PhotoMultiplier Tubes (MAPMT), each of 64 channels for a total of 2304 pixels with spatial resolution of $0.8^{\circ}$  
per pixel in timescales up to $2.5 \ \mu s$ called GTU (Gate Time Unit). 
Each MAPMT is powered by a Cockroft-Walton power 
supply board which convert 28V up to 1100V and present a BG3 UV filter on the entry window. The front end electronics consist of 6 SPACIROC3~\cite{Blin:2018tjp} (Spatial Photomultiplier 
Array Counting Integrated ReadOutChip) boards and 
a Xilinx Zynq XC7Z030 SoC board~\cite{belov2018}. In addition to the main detector, Mini-EUSO 
contains a Firefly MV and a Chameleon Point Grey compact cameras for complementary measurements in the near infrared and visible range, three single pixel UV sensors used 
as switches for day/night transition 
and a 64 channels Multi-Pixel Photon Counter (MPPC) imaging SiPM C13365 module provided by Hamamatsu Photonics. The front end electronic for SiPMs read out 
consist of a multiplexing board and an additional Atmel 2560 microcontroller board. 
The housekeeping of all instruments and the data storage into USB flash drive, is managed by a PCIe/104 
form factor CPU. The instrumentation is powered by the Low Voltage Power Supply and a custom power board. The LVPS  consisting of three PCB modules mounting different Vicor DC-DC 
converter which stabilize the $28$ V input voltage coming from ISS and provide power for all subsystems, preserving electronics from spike and polarization 
inversion. Primary and secondary grounds are galvanically isolated. The Mini-EUSO power consumption is around $55$ W. In Fig. \ref{int} the telescope hardware
and integration step.

\section{Space Qualification}

The General Technical Requirements for Experiment, Equipment and Technical Documents on board ISS required several tests to be performed on the instrument:
Electro-Magnetic Interference and Conductive (EMC/EMI); vibration and shock; high/low pressure, thermal and humidity functional tests, which are described in the following.
\subsection{EMC/EMI}
The aim of EMC/EMI tests is to verify that Mini-EUSO instrument does not produce any
undesired electromagnetic radiated emissions and is capable to withstand different
irradiations from external sources. In table \ref{emi}all EMI/EMC qualifying tests.

\begin{table}[h!]
\caption{\label{emi} EMI/EMC tests specifications.}
\begin{center}
\begin{tabular}{lll}

 \textbf{EMI/EMC}& \textbf{Frequency range ($KHz$)}& \textbf{Upeak ($dB\mu V$)}\\

Low-Frequency (LF) Conductivve Emissions&$0.03 \div 10$&$100\div 105$\\
High-Frequency (HF) Conductive Emissions&$0.009 \div 100$&$60\div 90$\\
Electric Field Produced by HF Emissions&$0.01 \div 1000$&$36\div 60$\\
Low-Frequency Conductive Interference&$0.02 \div 10$&$0.8\div 1$\\
Conductive HF Interference&$100 \div 3\times 10^{5}$&$91\div 120$\\

\end{tabular}
\end{center}
\end{table}

\begin{figure}[h!]
 \centering
 \includegraphics[scale=0.55,keepaspectratio=true]{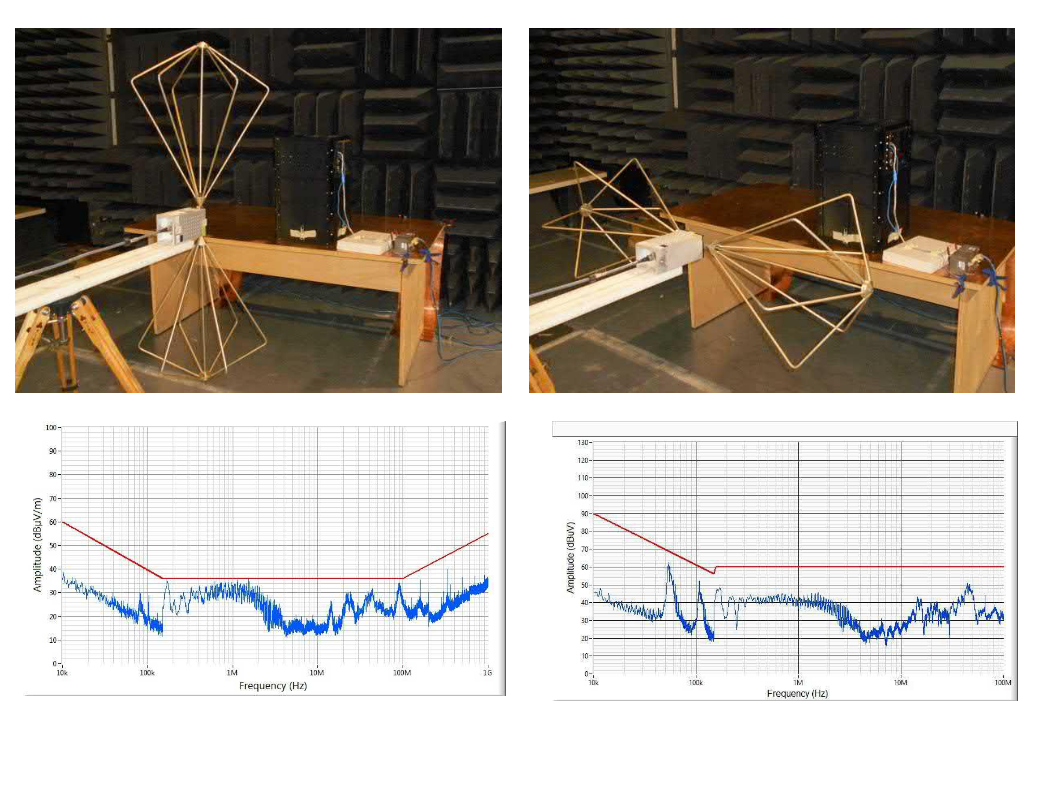}
 \caption{\label{emc}Top: Mini-EUSO Electric Field Interference tests (pol. V and H.). Bottom Left: E field Interference results.
 Bottom Right: HF conductive emissions results.}
\end{figure}

Mini-EUSO must not generate interference levels greater than the thresholds
specified when powered on both positive and negative wires at any 
supply voltages: 23, 28, 29 V. The detector was arranged in the anechoic 
chamber in vertical, in a way such that the
front end of the telescope faces the ground plane, 
with the front lens side touching the
table. This is due to the fact that during operations 
the telescope faces to the ISS window. 
Compliance with the thresholds has been assured for both horizontally and
vertically polarized waves using different types of antennas (Fig. \ref{emc}).
Generators were used to inject a voltage ripple profile into the input power lead and 
pulses with different amplitudes and durations. To comply
with the specifications for most of the circuit protection elements used in the onboard
hardware control system also Inrush current tests were
done. Start-up currents shall not exceed five times the maximum steady-state operating
current  during
power input voltage changes and pulses rise or fall rate should not exceed $0.125 A/\mu s$ with a maximum
energy of $0.1 \ A^2/s$. Mini-EUSO succesfully passed all EMC/EMI tests and the experience allowed the 
collaboration to improve electronics performances. In fig. \ref{emc} are shown 
the detector frequency profile responses, respectevely, when an external electric field is applied 
(left) and the high frequencies one conducted in normal operation (right).

\subsection{Temperature, Pressure and Humidity}
\begin{figure}[h!]
 \centering
 \includegraphics[scale=0.3,keepaspectratio=true]{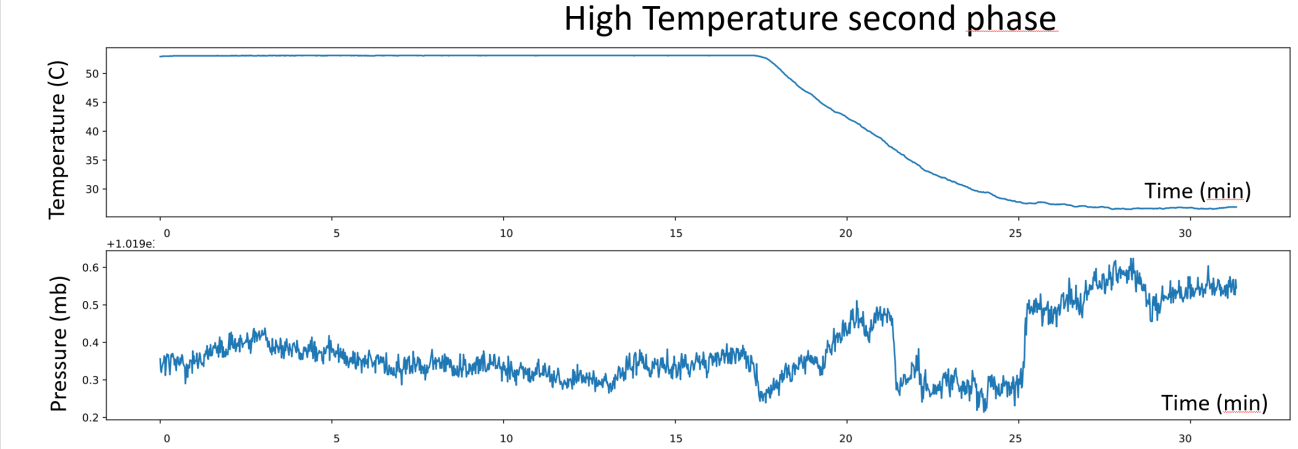}
 \caption{\label{temp1}High temperature cycle.}
\end{figure}

\begin{figure}[h!]
 \centering
 \includegraphics[scale=0.4,keepaspectratio=true]{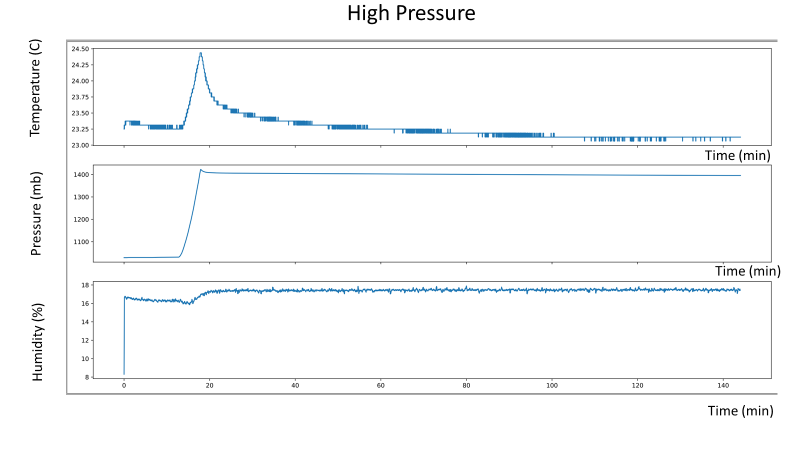}
 \caption{\label{temp2} High pressure cycle.}
\end{figure}

Pressure test can be divided into low and high pressure tests, each of them performed with pressure, temperature 
and humidity sensors read by a Raspberry Py Single Board Computer. Those requires the detector to be left inside a special chamber for two hours reaching pressures of 450 and 760 mm Hg 
in steps of 10 mm Hg per second. After reaching standard atmosphere Mini-EUSO normal functionality has to be checked.
Temperature test requires the detector to be installed into a thermal chamber for a total of 6 hours
in which it was kept respectevely at $\pm53^{\circ} C$. Check of nominal functionality was done after
reaching again standard temperature. The thermal cycles is shown in Fig. \ref{temp1}, while the pressure one
is visible in Fig. \ref{temp2}. Mini-EUSO electronics and optics have not been damaged at all and this is 
a milestone for future detector manufacturing.

\subsection{Vibration and Shock}
\begin{figure}[h!]
 \centering
 \includegraphics[scale=0.6,keepaspectratio=true]{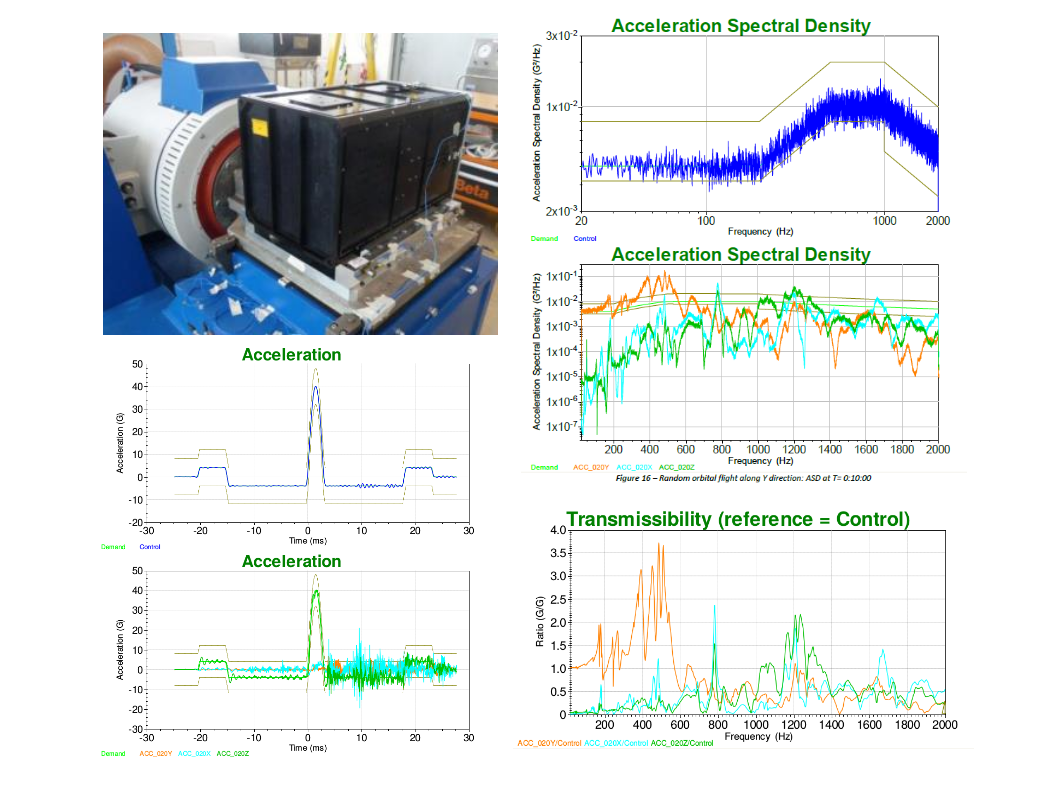}
 \caption{\label{shock} Top Left: Mini-EUSO accomodated on the vibrating plate. Bottom Left: Acceleration 
 response (X, Y and Z) during shock along Y axis. Right: Squared PSD for vibration along Y axis (top) 
 and overall trasmissibility (bottom).}
\end{figure}

Qualification vibration test requires random vibration and shock.
Tri-axial accelerometer for monitoring were located on the HW, while a mono-axial one was fixed to
the vibrating plate as feedback. A Resonance Survey was applied before and after each of the 
spectrums, where its level is defined as 
$0.5 sine - 2 octave/min - 1 \ sweep \ up \ only (20 \div 2000 Hz)$.
Random insertions were applied for 120, 480 and 600 seconds. The shock loads have a short time effect but 
with a high acceleration value ($\pm40 \ G$).
The two curves, before and after the tests, were compared. Each test is 
considered successfully completed if, after the visual
inspection, the equipment under test has maintained its physical characteristics, even if resonance
discrepancies exceed for more than $5 \%$. The detector resonances exceeded the thresholds in some tests 
even if no structural breaks has been exhibited and normal operations were succesful.
In Fig. \ref{shock} are shown plots of the resonance 
response after vibration and shock.

\section{Calibration} 
 
\begin{figure}[h!]
 \centering
 \includegraphics[scale=0.29,keepaspectratio=true]{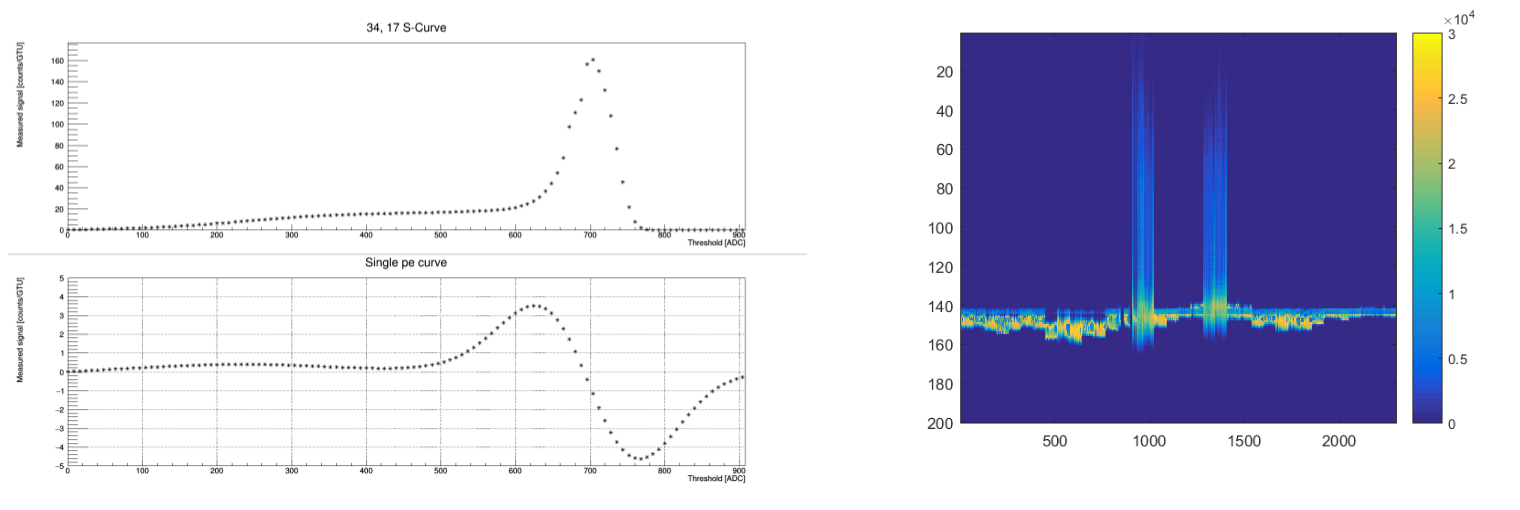}
 \caption{\label{scurve}Top Left: Single pixel S-curve. Bottom Left:
S-curve derivative.
In order to figure out
the threshold for the single PE production the
derivative must be analyzed. (On x-axis ADC threshold, on y-axis
measured counts). Right: Full PDM 2D scan
counts. On x-axis are shown the 2304
pixels as a function of ADC threshold
(y-axis). Only two ECs are powered
and this lead to the two trails in the
plot showing triggered pulses counts
(showed on the colour bar) also for low
ADC levels.}
\end{figure}

Mini-EUSO is intended to work in a single photoelectron (PE) counting mode which 
has advantages over analog measurement in terms of signal-to-noise ratio. The 64 signals from
MAPMT anodes are digitalized and discriminated to count photon
triggered pulses and to measure the photon intensity, thus, allowing to
set a threshold for the MAPMT single PE detection. This analysis is made
through the S-curve plot which showes the number of triggered pulses as a function of the ADC pulse height threshold. Each channels is an 8
bit threshold step, so that we can distinguish the typical noise
pedestal (low charge accumulation on the anode) from the lower rising slope representing photon reaching the focal surface (see Fig. \ref{scurve} left). 
It is possible also to adjust the gain noting that the pedestals, occurring around 150 ADC counts (see Fig. \ref{scurve} right), 
are shifted along different channels so the PE production is not uniform over all PDM. Note that the ADC bins are different in the two plot.  

\begin{figure}[h!]
 \centering
 \includegraphics[scale=2,keepaspectratio=true]{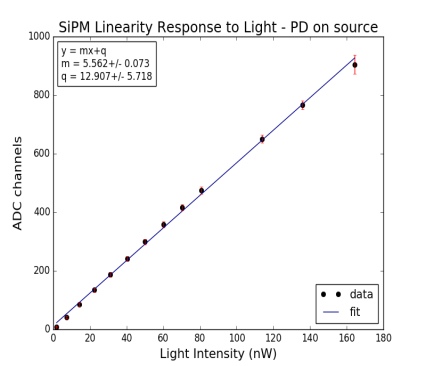}
 \caption{\label{sipm} SiPM ADC counts versus light intensity inside a black box.}
\end{figure}

Also for ancillary sensors an on-ground calibration was done.  
We chose $58 \ V$ power supply for the SiPM  C13365  array. The DC-DC converter provided by Hamamatsu was
programmed through a serial protocol and in Fig. \ref{sipm}, can be seen the range of its linear response to light of the SiPM Hamamatsu matrix.
For all optics calibration, sensors are located inside a black box. An integrating sphere focuses UV light emitted by diode onto the sensor's
entry window. Two calibrated photodiodes respectively arranged on the sphere and close to sensor collect photons which are converted in light 
intensity as:
\[
\dot{N_{ph}} =\frac{P_{light}}{E_{ph}}\simeq (2.014\times  10^9 \ ph/s)\times \bigl( \frac{P_{light}}{1 nW}\bigr) \times \bigl( \frac{\lambda}{400 nm}\bigr)
\]

\section{Conclusion}

Every test was positive and indeed the Mini-EUSO instrument on-board ISS is workin as foreseen. So far,
Mini-EUSO have been taking data for more than 40 sessions, more than a thousand of hours and it is still
running. This proves the stability and the capability of this kind of detector in space thus laying the foundations
for the future missions.



\end{document}